\documentclass[amsmath,amssymb,twocolumn,prl,showkeys,showpacs]{revtex4}

\usepackage{graphicx}
\usepackage[usenames]{color}

\begin{document}

\title{Pressure dependence of the charge-density-wave gap
in rare-earth tri-tellurides}

\author{A. Sacchetti$^{1}$, E. Arcangeletti$^{2}$, A. Perucchi$^{2}$, L.
Baldassarre$^{2}$, P. Postorino$^2$, S.
Lupi$^2$, N. Ru$^{3}$, I.R. Fisher$^{3}$, and L. Degiorgi$^{1}$}

\affiliation{$^{1}$Laboratorium f\"ur Festk\"orperphysik,
ETH-Z\"urich, CH-8093 Z\"urich, Switzerland. \\
$^{2}$CNR-INFM-Coherentia and Dipartimento di Fisica Universit\`a ``La
Sapienza'', P.le A. Moro 5, I-00185 Rome, Italy.\\
$^{3}$Geballe Laboratory for Advanced Materials and
Department of Applied Physics, Stanford University, Stanford,
California 94305-4045, USA.}

\date{\today}

\begin{abstract}
We investigate the pressure dependence of the optical properties of CeTe$_3$, which exhibits an incommensurate charge-density-wave
(CDW) state already at 300 K. Our data are collected in the
mid-infrared spectral range at room temperature and at pressures
between 0 and 9 GPa. The energy for the single particle excitation across
the CDW gap decreases upon increasing the applied pressure, similarly to the chemical pressure by
rare-earth substitution. The broadening of the
bands upon lattice compression removes the perfect nesting
condition of the Fermi surface and therefore diminishes the impact
of the CDW transition on the electronic properties of $R$Te$_3$.
\end{abstract}



\maketitle


The physical properties of low-dimensional systems have fascinated
researchers for a great part of the last century, and have
recently become one of the primary centers of interest in
condensed matter research. Low-dimensional systems not only
experience strong quantum and thermal fluctuations, but also admit
ordering tendencies which are difficult to realize in
three-dimensional materials. Prominent examples are spin- and
charge-density waves in  quasi-one-dimensional compounds
\cite{CDW}. Moreover, the competition among several possible order
parameters leads to rich phase diagrams, which can be tuned by
external variables as temperature, magnetic field, and both
chemical and applied pressure \cite{CDW,LeoBook}. Tunable external
parameters also affect the effective dimensionality of the
interacting electron gas, which plays an essential role in
defining the intrinsic electronic properties of the investigated
systems.

The rare-earth tri-tellurides $R$Te$_{3}$ ($R$= La-Tm, excepting
Eu \cite{Dimasi2}) are the latest paramount examples of low
dimensional systems exhibiting an incommensurate
charge-density-wave (CDW) state, stable across the available
rare-earth series \cite{Dimasi,struct}. The lattice constant decreases on going
from $R=$ La to $R=$ Tm \cite{footnote,LattConst}, i.e. by
chemically compressing the lattice, as consequence of the reduced
ionic radius of the rare-earth atom. The CDW state in $R$Te$_3$
can be then investigated as a function of the in-plane lattice
constant $a$, which is directly related to the Te-Te distance in
the Te-layers.

Recently, we have reported on the first optical measurements of
$R$Te$_3$ \cite{sacchetticond-mat}. Our data, collected over an
extremely broad spectral range, allowed us to observe both the
Drude component and the single-particle peak, ascribed to the
contributions due to the free charge carriers and to the
excitation across the charge-density-wave gap, respectively. We
established a diminishing impact of the charge-density-wave
condensate on the electronic properties of $R$Te$_3$ with
decreasing $a$ across the rare-earth series
\cite{sacchetticond-mat}. On decreasing $a$, a reduction of the
CDW gap together with an enhancement of the metallic (Drude)
contribution were observed in the absorption spectrum. This is the
consequence of a quenching of the nesting condition, driven by the
modification of the Fermi surface (FS) because of the lattice
compression \cite{sacchetticond-mat}.

We present in this letter infrared optical investigations of the pressure dependence of the optical reflectivity on
CeTe$_3$ at
300 K, i.e., below the CDW transition temperature. The motivation of this work originates from the fact
that $R$Te$_3$ generally provides an adequate playground to study
the effect of chemical pressure and externally applied pressure in
shaping the predisposition of these materials to undergo a CDW
phase transition. Upon increasing pressure the excitation due to
the CDW gap decreases in a quite equivalent manner when
compressing the lattice by substituting large with small ionic
radius rare-earth elements (i.e., by reducing $a$). These results
demonstrate that chemical and applied pressure similarly affect
the electronic properties and equivalently govern the onset of the
CDW state in $R$Te$_3$.

Single crystals of CeTe$_3$ were grown by slow cooling a binary
melt, as described elsewhere \cite{Ru}. A small piece of CeTe$_3$
(i.e., $50 \times 50$ $\mu$m$^2$) was cut from the same sample
previously used in Ref. \onlinecite{sacchetticond-mat} and was
placed on the top surface of a KBr pellet pre-sintered in the
gasket hole. The gasket was made of stainless steel, 50 $\mu$m
thick and with a 200 $\mu$m diameter hole. A clamp-screw diamond
anvil cell (DAC) equipped with high-quality type IIa diamonds (400
$\mu$m culet diameter) was employed for generating high-pressure up to 9 GPa.
Pressure was measured with the standard ruby-fluorescence
technique \cite{Mao}. Due to the metallic character of the sample,
absorption measurements are not possible on this compound.
Therefore, we carried out optical reflectivity measurements
exploiting the high brilliance of the SISSI infrared beamline at
ELETTRA synchrotron in Trieste \cite{SISSI}. The incident and reflected light
were focused and collected by a cassegrainian-based optical
microscope equipped with a MCT detector and coupled to a Bruker
Michelson interferometer, which allows to explore the 600-8000
cm$^{-1}$ spectral range. At each pressure, we measured the
light intensity reflected by the sample $I_S(\omega)$ and by
the external face of the diamond window $I_D(\omega)$, thus
obtaining the quantity
$R^S_D(\omega)=I_S(\omega)/I_D(\omega)$.
At the end of the pressure run, we also measured the light intensity reflected by a gold mirror ($I_{Au}(\omega)$) placed between the
diamonds at zero pressure and again $I_D(\omega)$, acting as a reference. One achieves $R_D^{\textrm{Au}}(\omega)=I_{Au}(\omega)/I_D(\omega)$, which is assumed to be pressure independent. This
procedure allows us to finally obtain the sample
reflectivity $R(\omega)=R^S_D(\omega)/R_D^{\textrm{Au}}(\omega)$ at each pressure, which takes into account the variations in the light intensity due to the smooth depletion of the current in the storage ring. The strong
diamond absorption at about 2000 cm$^{-1}$ and the presence of diffraction effects (see
below) prevent data reliability at low frequencies. Therefore, we
display the data in the 2700-8000 cm$^{-1}$ range.

\begin{figure}[!tb]
\center
\includegraphics[width=8.2cm]{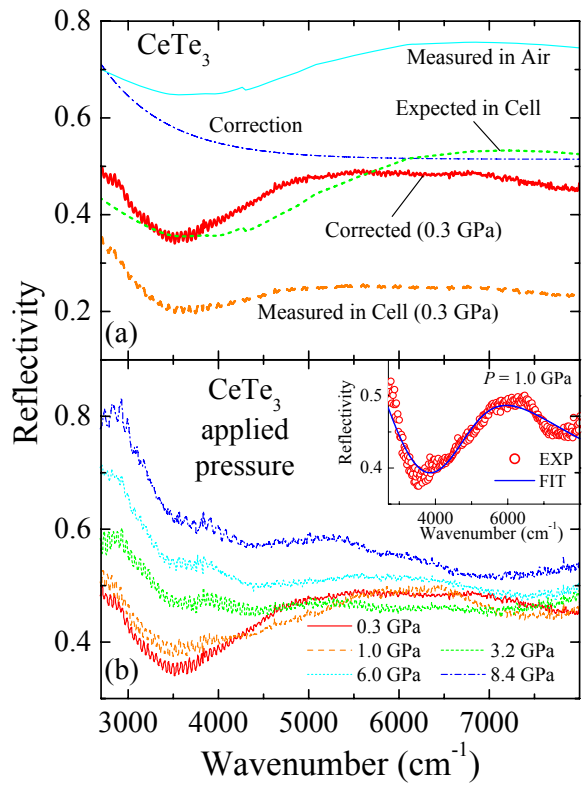}
\caption{(color online) (a) Raw $R(\omega)$ data of CeTe$_3$ at 300 K and 0.3 GPa
compared with the measured spectrum in air
\cite{sacchetticond-mat} and its expectation inside the DAC. The
correction function accounting for the diffraction effects is
reproduced, as well as the corrected reflectivity of CeTe$_3$ at 0.3 GPa. (b) $R(\omega)$ of CeTe$_3$ at 300 K and selected
applied pressures. The inset shows the very good reproducibility
of the data at 1 GPa within the Lorentz-Drude model (see text).}
\label{Refl}
\end{figure}

$R(\omega)$ of CeTe$_3$ at 300 K and 0.3 GPa is shown in Fig. 1a,
together with the corresponding $R(\omega)$ at ambient pressure
(i.e., outside the cell) \cite{sacchetticond-mat}.
Figure 1a also reproduces the expected $R(\omega)$ of CeTe$_3$
calculated from the complex
refractive index at zero pressure \cite{sacchetticond-mat} and assuming the sample inside the
DAC \cite{Wooten,dressel,simulation}. We immediately observe that the
expected $R(\omega)$ spectrum inside the DAC is lower than the one
at 0 GPa in air but still considerably higher than the
experimental finding. We ascribe this difference to diffraction
effects induced by the non-perfectly-flat shape of the sample. In
order to take into account these diffraction effects, we define a
smooth correction function (Fig. 1a) which is then applied to all spectra. We justify our choice for the correction
function by pointing out that it is somehow more effective at high frequencies (as expected
for diffraction effects) and that it shows a strong frequency dependence only below
2700 cm$^{-1}$. Furthermore, we checked that the final corrected spectra as well as the data analysis do not substantially change
when correction
procedures based on a simple scaling by a constant factor or on adding a constant background to the measured spectra are employed.   

Figure 1b reproduces the corrected spectra of CeTe$_3$ at selected
pressures. Although the light spot was precisely limited (by means
of fissures) to the sample area, there is still some diffused
light giving rise to the interference pattern (between the diamond
windows) observed in the spectra. The striking feature is the
filling-in of the deep minimum in $R(\omega)$ at about 3500
cm$^{-1}$ with increasing pressure, quite similar to the behavior
of $R(\omega)$ across the rare-earth series (inset of Fig. 1 in
Ref. \onlinecite{sacchetticond-mat}).
\begin{figure}[!tb]
\center
\includegraphics[width=8.2cm]{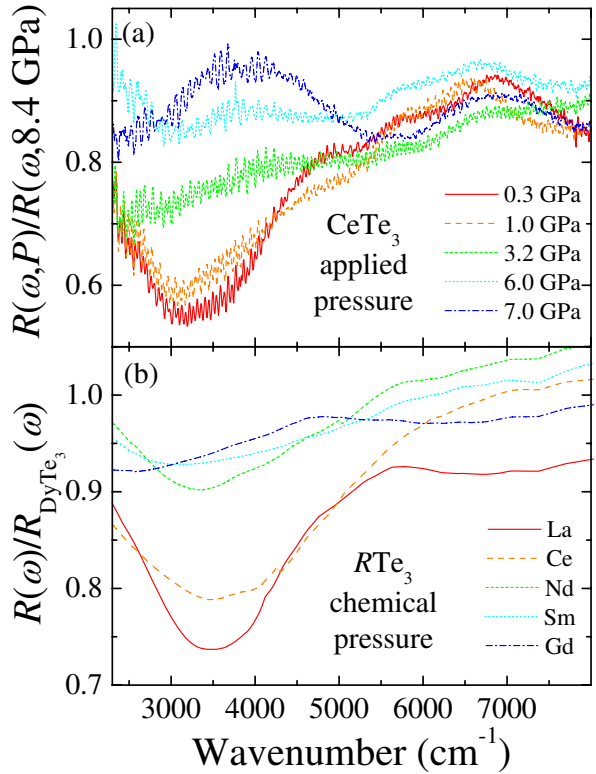}
\caption{(color online) (a) $R(\omega)$ of CeTe$_3$ at 300 K and selected
applied pressures, normalized by the spectrum at 8.4 GPa. (b)
$R(\omega)$ of $R$Te$_3$ at 300 K and ambient pressure, normalized
by the spectrum of DyTe$_3$ \cite{sacchetticond-mat}. } \label{ratioRefl}
\end{figure}
The depletion at 3500 cm$^{-1}$ was ascribed to the charge
excitation across the CDW gap into a single particle ($SP$) state
\cite{sacchetticond-mat}. A more compelling comparison is given in
Fig. 2, displaying the ratio of the $R(\omega)$ spectra of
CeTe$_3$ at selected pressures with respect to the spectrum at the
highest measured pressure (Fig. 2a) and the ratio of the
$R(\omega)$ spectra for selected rare-earth compounds with respect
to $R(\omega)$ of DyTe$_3$ (Fig. 2b) \cite{sacchetticond-mat}. The obvious similarity
between the $R(\omega)$ ratios upon increasing pressure and when
moving from the La to the Dy compound suggests the equivalence
between chemical and applied pressure.

The optical findings on the rare-earth series were systematically
reproduced within the Drude-Lorentz fit: the most relevant
components were the Drude term ascribed to
the effective metallic contribution and three Lorentz harmonic
oscillators (h.o.) ascribed to the $SP$ excitation. The same fit
procedure (Fig. 3a) is applied here to the pressure dependent $R(\omega)$
spectra of CeTe$_3$ \cite{simulation}. Only the parameters of the
three Lorentz h.o.'s, describing the $SP$ excitation, were allowed
to change as a function of pressure. All other components (Drude
term and electronic interband transitions) were left fixed, by
exploiting the best fit of the CeTe$_3$ data at ambient pressure
outside the DAC \cite{sacchetticond-mat}. We also tested slightly
different fitting procedures obtaining similar results. By fitting
$R(\omega)$ spectra in the energy interval displayed in Fig. 1b,
we can then reconstruct the real part $\sigma_1(\omega)$ of the
optical conductivity of CeTe$_3$ at selected pressures. This is
shown in Fig. 3a, while the inset of Fig. 1b exemplifies the good
fit quality of $R(\omega)$ at 1 GPa. There is an overall good
correspondence with $\sigma_1(\omega)$ of the rare-earth series
(Fig. 3b), reproduced from Ref. \onlinecite{sacchetticond-mat}.
\begin{figure}[!tb]
\center
\includegraphics[width=8.2cm]{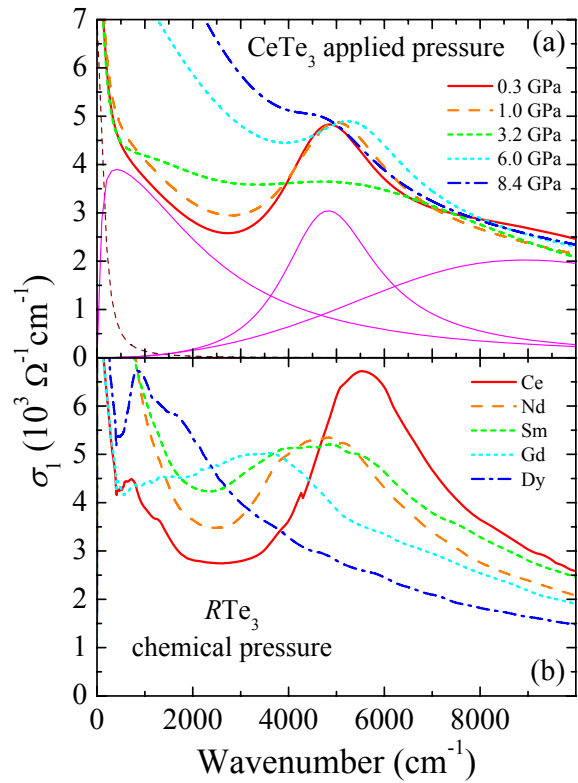}
\caption{(color online) (a) Real part $\sigma_1(\omega)$ of the optical
conductivity of CeTe$_3$ at 300 K and selected pressures,
calculated from the fit of $R(\omega)$ within the Lorentz-Drude
model (see text). The Drude-Lorentz fit components at 0.3 GPa are also displayed. (b) $\sigma_1(\omega)$ at 300 K of $R$Te$_3$
($R$=Ce, Nd, Sm, Gd, and Dy), obtained through Kramers-Kronig
transformation of the measured $R(\omega)$ spectra
\cite{sacchetticond-mat}.} \label{Sig1}
\end{figure}

In order to push further the comparison between chemical (i.e.,
rare-earth dependence) and applied pressure, we first need to
establish the pressure dependence of the lattice constant $a(P)$.
A direct experimental determination of $a(P)$ is still missing,
but we can extract this latter quantity from the zero-pressure
bulk modulus $B_0$. First of all, from the $\beta$-value $\biggl\lbrack$$\beta=\frac{2\pi}{5}k_B(\frac{k_B}{\hbar v_s})^3$$\biggl\rbrack$ of the
phononic part of the specific heat in LaTe$_3$
\cite{Ru,SpecHeat} one achieves the sound velocity $v_s=1923$ m/s
\cite{note}. Knowing that $B_0=\rho v_s^2$, $\rho$=6837 kg/m$^3$
being the density, one gets $B_0=25$ GPa. We can then assume a
linear pressure dependence of the bulk modulus $B(P)=B_0+B'P$,
where $B'$ usually ranges between 4 and 8 \cite{Bprime}. This
leads to the so-called Murnaghan equation for the pressure
dependence of the volume \cite{Murnaghan}:
\begin{equation}
 V(P)=V(0)\Big(1+\frac{B'}{B}P\Big)^{-1/B'},
\end{equation}
from which we can immediately obtain $a(P)=a(0)[V(P)/V(0)]^{1/3}$.
The inset of Fig. 4 shows the range within which the pressure
dependence of $a$ can evolve for the two limits of $B'$. 

The three h.o.'s for the $SP$ excitation (Fig. 3a) allow us to define the
so-called averaged excitation energy $\omega_{SP}$ (eq. (2) in
Ref. \onlinecite{sacchetticond-mat}) at each pressure. The main
panel of Fig. 4 displays the resulting
dependence of $\omega_{SP}$ on the lattice constant;
$\omega_{SP}(a)$ for CeTe$_3$ at different pressures is determined
for the average of $a(P)$ between the $B'$=4 and 8 curves (inset
of Fig. 4), while $\omega_{SP}(a)$ for the rare-earth series is
reproduced from Fig. 3b of Ref. \onlinecite{sacchetticond-mat}.
\begin{figure}[!tb]
\center
\includegraphics[width=8.2cm]{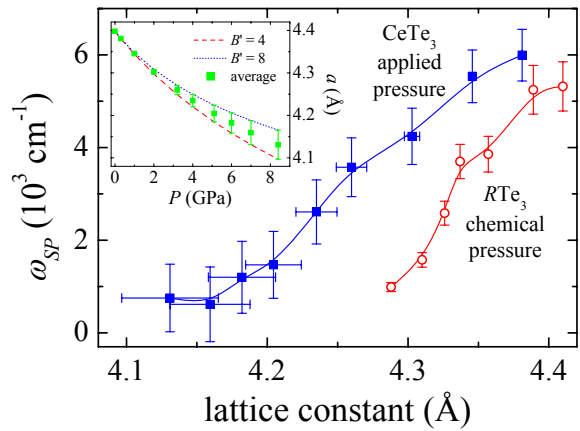}
\caption{(color online) Single particle excitation energy $\omega_{SP}$ as a
function of the lattice constant $a$ for CeTe$_3$ under applied pressures and
for the $R$Te$_3$ series \cite{sacchetticond-mat}. Solid lines are guides to the eye. Inset:
calculated pressure dependence of $a$ (see
text).} \label{gap-pressure}
\end{figure}
There is again a similar trend between the two sets of data, even
though the two curves do not fully overlap. A perfect
correspondence is anyhow not expected in view of the
approximations, employed for the determination of $a(P)$. In
particular, the observed discrepancy could be ascribed to an
underestimate of $B_0$, as well as to the assumption of isotropic
compression ($a(P) \propto V^{1/3}(P)$). It appears that
$\omega_{SP}$ gets smaller upon decreasing $a$. Analogous to the
rare-earth series \cite{sacchetticond-mat}, such a reduction of
$\omega_{SP}$ on decreasing $a$ may be considered as an indication
for the lesser impact of the CDW state upon increasing pressure.
Pressure changes the shape of FS in such a way to alter the
favorable nesting conditions, which are the prerequisite for the
formation of the CDW condensate \cite{CDW}. Lattice compression
broadens the bands so that the amount of the nested FS diminishes,
as well. This favors a shift of spectral weight from the $SP$ peak
to low frequencies and induces the filling-in of the CDW gap
feature in the excitation spectrum (Fig. 3)
\cite{sacchetticond-mat}. There is an indirect support to these
conclusions by a recent angle resolved photoemission spectroscopy
experiment \cite{ARPES4}, where a reduction of the CDW gap with
decreasing $a$ was observed for several compounds of the $R$Te$_3$
series.

In conclusion, we have reported the first optical investigation of
the pressure dependence of the single particle excitation across
the CDW gap in CeTe$_3$. Pressure affects the gapping of FS so
that the CDW gap is progressively suppressed on decreasing the
lattice constant. Therefore, our findings confirm the
equivalence between applied and chemical pressure in the
rare-earth tri-telluride series. The formation of the CDW state in
$R$Te$_3$ was also considered as an indication for a hidden
one-dimensional behavior in these quasi two-dimensional compounds
\cite{sacchetticond-mat,Kivelson}. This work does not address to
which extent the applied pressure might influence the effect of
electron-electron interactions and Umklapp processes, as suggested in Ref.
\onlinecite{sacchetticond-mat}, as well as the dimensionality crossover,
in driving the CDW transition. This awaits for further
experimental effort, allowing the extension of the measured
spectral range under pressure up to higher as well as to lower
energies than the energy window presented here. This could open
new perspectives to a comprehensive study about the pressure
dependence of the characteristic power law behavior, seen in the
absorption spectrum of the $R$Te$_3$ series
\cite{sacchetticond-mat}, and more generally about the influence
of pressure in the formation of the Luttinger liquid state in
quasi one-dimensional systems.

The authors wish to thank D. Basov and T. Giamarchi for fruitful
discussions. One of us (A.S.) wishes to acknowledge the
scholarship of the Della Riccia Foundation. This work has been
supported by the Swiss National Foundation for the Scientific
Research within the NCCR MaNEP pool and also by the Department of
Energy, Office of Basic Energy Sciences under contract
DE-AC02-76SF00515.


\begin{thebibliography}{99}


\bibitem{CDW} G. Gr\"uner, \emph{Density Waves in Solids},
Addison Wesley, Reading, MA (1994).

\bibitem{LeoBook} \emph{Strong Interactions in Low Dimensions},
Eds. D. Baeriswyl and L. Degiorgi, Kluwer Academic Publishers,
Dordrecht (2004).

\bibitem{Dimasi2} E. DiMasi, B. Foran, M.C. Aronson, and S. Lee,
\emph{Chem. Mat.} \textbf{6}, 1867 (1994).

\bibitem{Dimasi} E. DiMasi \textit{et al.}, \emph{Phys. Rev. B}
\textbf{52}, 14516 (1995).

\bibitem{struct} B.K. Norling and H. Steinfink, \emph{Inorg.
Chem.} \textbf{5}, 1488 (1966).

\bibitem{footnote} The crystal structure of $R$Te$_3$ belongs to
the $Cmcm$ space group, which is orthorhombic. However, the
in-plane lattice constants $a$ and $c$ differ by approximately
0.1\% (in the standard space group setting, the $b$-axis is
perpendicular to the Te planes). For simplicity, in our subsequent
analysis we treat the material as essentially tetragonal,
characterized by an in-plane lattice constant $a$.

\bibitem{LattConst} P. Villars and L.D. Calvert, \emph{Pearson's
Handbook of Crystallographic Data for Intermetallic Phases},
American Society for Metals, Metals Park, OH (1991).

\bibitem{sacchetticond-mat} A. Sacchetti \textit{et al.}, \emph{Phys. Rev. B} (in press) and \emph{cond-mat/0606451}.

\bibitem{Ru} N. Ru and I.R. Fisher, \emph{Phys. Rev. B}
\textbf{73}, 033101 (2006).

\bibitem{Mao} H.K. Mao, J. Xu, and P.M. Bell, \textit{J. Geophys.
Res.} \textbf{91}, 4673 (1986).

\bibitem{SISSI} S. Lupi \textit{et al.}, submitted to \emph{J. Opt. Soc. Am. B}.

\bibitem{Wooten} F. Wooten, \emph{Optical Properties of Solids},
Academic Press, New York (1972).

\bibitem{dressel} M. Dressel, and G. Gr\"uner, {\itshape
Electrodynamics of Solids}, Cambridge University Press (2002).

\bibitem{simulation} The reflection coefficient for the experimental arrangement given by the sample and the diamond window of the DAC is defined by $\hat{r}=(n'-\hat{n})/(n'+\hat{n})$
\cite{dressel}, $n'=2.42$ being the refractive index of diamond
$\lbrack$P. Dore \textit{et al.}, \textit{Appl. Optics}
\textbf{37}, 5731 (1998)$\rbrack$ and $\hat{n}$ the complex refractive index of
CeTe$_3$ \cite{sacchetticond-mat}.

\bibitem{SpecHeat} K.Y. Shin \textit{et al.}, \emph{Phys. Rev. B} \textbf{72}, 085132 (2005).

\bibitem{note} For LaTe$_3$ $\beta_{meas}=1.33 \times 10^{-3}$ J
mol$^{-1}$K$^{-4}$ \cite{SpecHeat}. Since
$\beta=\beta_{meas}/V_m$, where the molar volume is $V_m=N_A a^2
c/4=7.63 \times 10^{-5}$ m$^3$mol$^{-1}$ (the unit cell contains 4 f.u.), we obtain $\beta=17.4$ J
m$^{-3}$K$^{-4}$.

\bibitem{Bprime} S. Jiuxun \textit{et al.}, \textit{J. Phys. Chem.
Solids} \textbf{66}, 773 (2005).

\bibitem{Murnaghan} F.D. Murnaghan, \textit{P. Natl. Acad. Sci. U.S.A.}
\textbf{30}, 244 (1944).

\bibitem{ARPES4} V. Brouet \emph{et al.}, private communication.

\bibitem{Kivelson} H. Yao \textit{et al.}, \emph{cond-mat/0606304}.

\end{thebibliography}
\end{document}